\documentclass[a4paper,american,floatfix,pdftex,superscriptaddress,twoside,%
aip,pof,%
citeautoscript,
 preprint,%
reprint,twocolumn
]{revtex4-2}

\usepackage{caption}
\usepackage{subcaption}
\usepackage{graphicx}
\usepackage{amsmath}
\usepackage{tabularx}

\usepackage{color}
\usepackage{xspace,commath}
\usepackage[pdftex,pdfpagelabels,plainpages=false,linktocpage=true]{hyperref}
\hypersetup{colorlinks=true,citecolor=blue,urlcolor=blue,filecolor=green,
                    linkcolor=red}

\newcommand{\Kn}{\ensuremath{\operatorname{\mathit{K\kern-.20em n}}}\xspace}
\newcommand{\cmiweb}{\homepage[website:~]{https://www.controlled-molecule-imaging.org}}%

\DeclareGraphicsExtensions{.eps,.pdf,.png}
\graphicspath{{./pictures/}} 

\begin{document}


\title{An improved numerical simulation methodology
for nanoparticle injection through aerodynamic
lens systems}



\author{Surya Kiran Peravali}
\email[]{peravals@hsu-hh.de}
\cmiweb
\affiliation{Professur für Strömungsmechanik, Helmut-Schmidt-Universität / Universität der Bundeswehr Hamburg, 22043 Hamburg, Germany}
\affiliation{Center for Free-Electron Laser Science CFEL, Deutsches
      Elektronen-Synchrotron DESY, Notkestr.~85, 22607 Hamburg, Germany}

\author{Amit K.\ Samanta}
\affiliation{Center for Free-Electron Laser Science CFEL, Deutsches
      Elektronen-Synchrotron DESY, Notkestr.~85, 22607 Hamburg, Germany}
\affiliation{Center for Ultrafast Imaging, Universität Hamburg, Luruper
      Chaussee 149, 22761 Hamburg, Germany}

\author{Muhamed Amin}
\affiliation{Center for Free-Electron Laser Science CFEL, Deutsches
      Elektronen-Synchrotron DESY, Notkestr.~85, 22607 Hamburg, Germany}
\affiliation{Laboratory of Computational Biology, National Heart, Lung
             and Blood Institute, National Institutes of Health, Bethesda, 
             Maryland 20892, USA}

\author{Philipp Neumann}
\affiliation{Department of Informatics, High Performance Computing \& Data Science, Universität Hamburg, 22603 Hamburg, Germany}
\affiliation{Deutsches Elektronen-Synchrotron DESY, Notkestr.~85, 22607 Hamburg,
      Germany}

\author{Jochen Küpper}
\affiliation{Center for Free-Electron Laser Science CFEL, Deutsches
      Elektronen-Synchrotron DESY, Notkestr.~85, 22607 Hamburg, Germany}
\affiliation{Center for Ultrafast Imaging, Universität Hamburg, Luruper
      Chaussee 149, 22761 Hamburg, Germany}
\affiliation{Department of Physics, Universität Hamburg, Luruper Chaussee 149,
      22761 Hamburg, Germany}

\author{Michael Breuer}
\email[]{breuer@hsu-hh.de}
\affiliation{Professur für Strömungsmechanik, Helmut-Schmidt-Universität / Universität der Bundeswehr Hamburg, 22043 Hamburg, Germany}

%

\date{\today}

\begin{abstract}
Aerosol injectors applied in single-particle diffractive imaging experiments demonstrated their potential in efficiently delivering nanoparticles with high density. Continuous optimization of injector design is crucial for achieving high-density particle streams, minimizing background gas, enhancing X-ray interactions, and generating high-quality diffraction patterns. We present an updated simulation framework designed for the fast and effective exploration of the experimental parameter space to enhance the optimization process. The framework includes both the simulation of the carrier gas and the particle trajectories within injectors and their expansion into the experimental vacuum chamber. A hybrid molecular-continuum-simulation method (DSMC/CFD) is utilized to accurately capture the multi-scale nature of the flow. The simulation setup, initial benchmark results of the coupled approach, and the validation of the entire methodology against experimental data are presented.
\end{abstract}


\maketitle 


\section{Introduction}

\noindent Single-particle diffractive imaging (SPI) is a novel technique used for imaging atomic-scale structures ranging from few micrometers to nanometers~\cite{SPI1, SPI2}, such as bio-molecules, proteins or artificial nanoparticles. In this approach identical particles are delivered, in a high-density stream, into vacuum where they are intersected with X-ray free-electron laser (XFEL) pulses. When an X-ray pulse hits the particle in flight, a two-dimensional diffraction pattern is produced. Collecting a large set of such diffraction patterns of identical particles, allows for the reconstruction of the particles three-dimensional structure~\cite{Ekeberg2015, Seibert2011, Ayyer:Optica7:593}. Particle that interacted with the intense X-ray pulses are destroyed. Therefore, a continuous stream of identical particles is required, which can be achieved using aerosol injectors. Aerodynamic-lens-stack (ALS) injectors are most commonly used at XFEL facilities to provide focused or collimated nanoparticle beams for SPI experiments~\cite{Ekeberg2015,ROTH2024}. 

An ALS consists of a series of orifices, traversed by the particles in the gas phase, from which a particle stream is extracted into vacuum. The typical design and setup of an ALS in an SPI experiment were described elsewhere~\cite{Worbs:te5083}. The sample-injection system must be optimized in order to produce high-quality particle beams, i.e., a high particle density to increase the hit rate with the X-ray pulse and a low carrier gas density to reduce background scattering~\cite{Worbs:te5083}. The latter necessitates shifting the particle-beam focus away from the ALS exit. Optimizing the ALS design based on experimental characterization in its large parameter space is time consuming and thus often impractical. A computational approach can serve as a fast and efficient alternative to investigate the parameters, e.g., flow rate, pressure, and carrier gas, that control the particle-beam size and the focusing behavior. 

Single and multi-lens systems for particle beam collimation by an ALS were characterized utilizing numerical simulations~\cite{Zhang1,Zhang2}. A numerical study described the focusing of particles to a beam with a diameter smaller than 30~nm using an ALS~\cite{Wang1, Wang2}. This work established the guidelines for designing aerodynamic lens systems for nanoparticles and also a design tool that predicts ALS dimensions to focus particles of certain sizes at different flow conditions. However, in all these studies, the flow through the ALS was assumed to be a \emph{continuum} as numerical solvers based on continuum mechanics, i.e., the Navier-Stokes equations, were used to predict the gas flows. Particle trajectories were computed based on forces determined from these simulated flow fields. This numerical methodology was adopted for simulating nanoparticle-injection experiments at XFEL facilities~\cite{ROTH201817}, which further led to the development of an in-house particle trajectory simulation tool denoted CMInject~\cite{welker2022}. Here, the drag force model used for calculating the particle movement in the fluid is described by Stokes' law. While the flow field is in the continuum range in these specific cases, non-continuum effects prevail with respect to the small particles as in certain cases the particle diameters can be smaller than the mean-free path of the fluid (Knudsen number $\Kn_p>1$) leading to decreased drag forces. To take this into account,
the empirical Cunningham slip-correction factor~\cite{cunningham1910} was used along with the Stokes drag. Additionally, a Brownian-motion force was added to the drag term in order to incorporate the Brownian motion of the nanoparticles.

The nanoparticle focusing behavior in multi-scale flow regimes, i.e., transition and free-molecular
flow regime, is largely unexplored~\cite{Wang1}. In these regimes, the flow Knudsen number $\Kn$
has a larger value ($\Kn>0.1$) such that the continuum assumption for the flow is violated.
Therefore, traditional Navier-Stokes-based CFD solvers fail to accurately resolve the flow and
particle-based Boltzmann solvers, such as direct simulation Monte Carlo (DSMC), have become the
method of choice. However, the DSMC method is challenging if the simulation includes both continuum
and rarefied regions. Furthermore, this method is computationally very inefficient for small Knudsen
numbers ($\Kn<0.1$). This necessitates the use of a hybrid approach combining DSMC with CFD.

The Stokes-Cunningham drag model described and used in previous numerical works~\cite{Zhang1, Zhang2, Wang1, Wang2, ROTH201817, welker2022} 
is confined to continuum gas flow fields at low Mach numbers and also strictly depends on empirical relations. For rarefied flow regimes ($\Kn_p > 1$), the drag force on spherical particles at small Mach numbers was described by the Epstein model~\cite{Epstein}. Unlike the Cunningham model, which assumes the gas molecules to be specularly reflected on the surface of the sphere, the Epstein drag model assumes a combination of both specular and diffusive gas-surface collisions. A closed-form expression for the drag force on small spheres in the free-molecular regime for all Mach numbers was described~\cite{Baines}. Furthermore, several studies reported on the generalization of the drag force model to encompass a broad spectrum of Reynolds and Mach numbers. These works relied on either ad-hoc interpolations between different regimes~\cite{henderson1976,Loth2008}, empirical correlations from the available literature~\cite{DAVULURI2021} or neural-network based empirical formulations~\cite{LI2019-NNdrag}. In recent years, a derivation of a generalized physics-based expression for the drag coefficient of spherical particles was attempted~\cite{Nsingh_Tom2021}. For highly rarefied regimes where gas can tend toward non-equilibrium, a DSMC based approach for computing force on a particle was introduced~\cite{Gallis_DSMC_drag,burt_phd2006}. This is advantageous where the molecular distribution of the gas is not known beforehand and can only be determined through DSMC computations. The main disadvantage of this model are that it can be inaccurate in the low Knudsen number regime and that it is computationally inefficient.

We present a new, improved simulation framework utilizing a one-way coupled CFD-DSMC methodology to
resolve the gas flow through ALS systems in the presence of different Knudsen number regimes. This
hybrid CFD-DSMC methodology was already validated~\cite{peravali2024accuracy} based on a gas-dynamic
nozzle case~\cite{rothe}. Here, the motion of the nanoparticles through the multi-scale ALS flow is
modeled and the particle interaction with the background gas is examined. The accuracy of the entire
simulation tool is evaluated by comparing the simulations with experimental
data~\cite{worbs_phd2022}. Finally, the nanoparticle-focusing behavior through the ALS is studied in
detail based on the improved simulation framework along with additional corrections of the molecular
drag models~\cite{Epstein, Baines} at extremely rarefied flow conditions.

\section{Test case and experimentation}
\label{sec:testcase}

\begin{figure}
    \centering
    \includegraphics[width=0.9\linewidth]{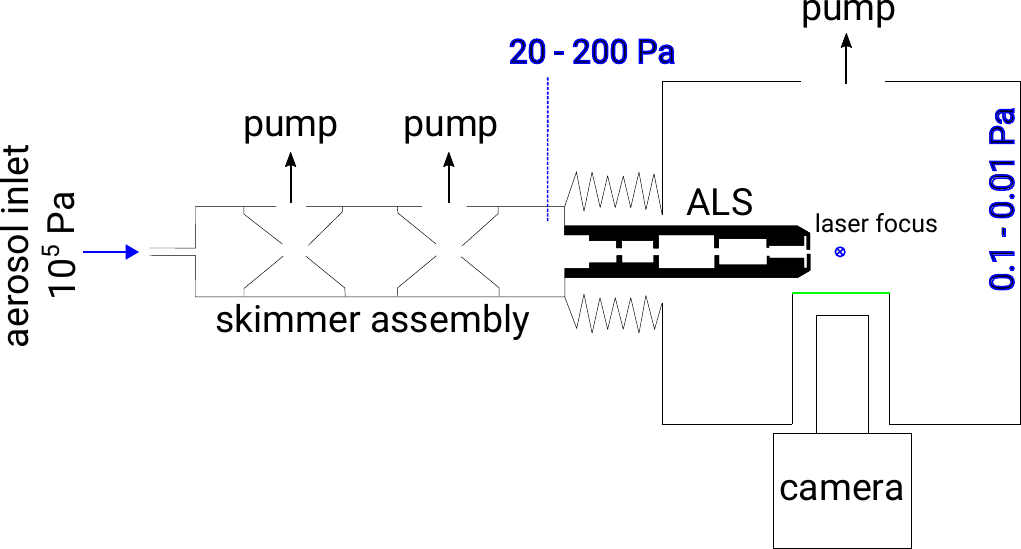}
    \caption{Schematic of a typical experimental setup used in the particle-beam evolution
      measurements~\cite{worbs_phd2022}. The setup consists of a double skimmer setup with adjustable pumping, an
      aerodynamic-lens-stack assembly for particle-beam generation, and the optical scattering setup and particle-localization microscopy inside a high-vacuum chamber.}
    \label{fig:ALS_experiment}
\end{figure}

The nanoparticle beams were generated using the experimental setup shown in~\autoref{fig:ALS_experiment}. Polystyrene spheres (PS) were aerosolized at pressures of about $10^5$~Pa and passed through a differentially-pumped skimmer assembly to reduce gas-load and pressure in the experiment. The particles were focused, through an ALS, into the ultra-high vacuum detection chamber ($p \approx 10^{-1}$~Pa). These experiments were carried out for different ALS inlet-gas pressures  ($p_{in}$) and different particle sizes summarized in~\autoref{tab:flowcases} along with the flow Knudsen numbers ($\Kn$) and particle Knudsen numbers ($\Kn_p$) at the inlet of the ALS. 
The particle beam profiles were obtained through particle-localization microscopy~\cite{Awel:16} at different distances from the exit of the ALS. Detailed descriptions of the ALS geometry used in the experiment, the experimental procedure, and the analysis of the experimental data were described elsewhere~\cite{worbs_phd2022}. 

\begin{table}[h!]
	\centering
	\color{black}
	\begin{tabular}{l@{\extracolsep{3ex}} c c c}
		\hline
		\textbf{Particle size} & Inlet pressure, $p_{in}$ &$\Kn$ &${\Kn_p}$ \\
		\hline
		69~nm& 180~Pa& 0.0241 & 523.87 \\
		
		    & 55~Pa  & 0.0788   & 1714.5\\
		    & 20~Pa & 0.2168   & 4714.88 \\
      \hline
		42~nm& 180~Pa& 0.0241 & 860.65 \\
		
		    & 55~Pa  & 0.0788   & 2816.68 \\
		    & 20~Pa  & 0.2168   & 7745.87  \\
		\hline

        25~nm& 200~Pa& 0.0213& 1279.196 \\
		
		    & 150~Pa  & 0.0289   & 1735.075 \\
		    & 50~Pa  & 0.086   & 5205.225 \\
		\hline
	\end{tabular}
	\caption{\label{tab:flowcases} Gas-flow and particle parameters
                                       of the experiments.}
\end{table}

\section{Simulation methodology}
\label{sec:method}

In the described experiments, the volume fraction of the particles in the gas-particle mixture was very low. Therefore, it is assumed that the particles do not influence the flow field of the gas and that there is no, or only negligible interaction between particles. This implies that the gas-particle dynamics can be calculated in a decoupled manner and we employ a two-step approach to calculate the gas-particle dynamics: First, we calculate the flow field through the ALS, which is converged to a steady-state solution. Second, the particles are tracked independently by interpolating the forces obtained from the flow field. We utilize various methods for resolving the fluid field in different regimes, which are described in the following subsections. Furthermore, different models for the forces that influence the particle transport are described.

\subsection{Flow field in continuum (CFD)}
\label{subsec:continuumflow}

For the experiments with higher inlet pressures, i.e., when the Knudsen numbers $\Kn$ of the flow field throughout the ALS and near its exit are small ($\Kn < 0.1$) the flow can be described as continuum ($0 < \Kn < 0.01 $) or in a slip regime ($0.01 < \Kn < 0.1$). For these regimes, the flow field can be computed by solving the Navier-Stokes equations. The continuum gas flow field is computed using the finite-volume software OpenFOAM~\cite{openfoam}. Since the flow through the ALS transits from subsonic to supersonic speeds in streamwise direction, the flow has to be assumed as compressible. A density-based transient solver (\texttt{rhoCentralFoam}) is utilized. Detailed information on the solver settings, e.g., discretization, interpolation and boundary conditions can be found in a previous work~\cite{peravali2024accuracy}. Since in this case the Reynolds number is very low ($Re < 10$), the flow is assumed to be laminar and the equation of state for a perfect gas was applied. The transport properties are estimated using the Sutherland transport model~\cite{white2006}. The CFD calculations of the test cases rely on the structured grid specified and depicted in~\appref{app:OF}.

\subsection{Highly rarefied flow (DSMC)}
\label{subsec:rarefiedflow}

The pressure at the inlet of the ALS is one of the major tuning factors of particle injection in SPI
experiments. Sometimes very low inlet pressures $p_{in}<50$~Pa are used as this reduces the
background X-ray scattering. For such low pressures, the gas flow corresponds to a larger mean free
path between gas molecules and the Knudsen number is larger than $0.1$. To resolve the flow field in
these regimes, the direct simulation Monte Carlo (DSMC) method is often a good
choice~\cite{Bird1994}. It is a stochastic technique, which provides an approximate solution to the
Boltzmann equation~\eqref{eq:boltmann}:
\begin{equation}\label{eq:boltmann}
    \frac{\partial f}{\partial t} + \textbf{u} \cdot \nabla f = \left (\frac{\partial f}{\partial t} \right )_\text{coll}  \; .
\end{equation}

Here, any external forces are assumed to be negligible. Each simulation particle represents a large
number of real gas molecules, maintaining the phase space of the overall distribution. The momentum
term ($\textbf{u} \cdot \nabla f $) and the collision term
$\left (\frac{\partial f}{\partial t} \right )_\text{coll}$ are solved in a decoupled manner. The
probabilistic models are utilized to solve the collision term and also the relaxation of internal
degrees of freedom. In this work, we used the DSMC software SPARTA (Stochastic PArallel Rarefied-gas
Time-accurate Analyzer)~\cite{SPARTA}. The DSMC solution is sensitive to several parameters, such as
the number of simulation particles, the grid size of the computational domain, the time step, the
inter-molecular/surface collision models, and the sampling. The ideal choice of these parameters
depends on various other factors like operating conditions, Reynolds number and gas/mixture
properties, etc. A variety of such options in SPARTA were evaluated in terms of the accuracy and
performance of the solution~\cite{peravali2024accuracy}, establishing guidelines for accurate and
efficient DSMC simulations, which are incorporated in the present study and noted in
\appref{app:dsmc}.

\subsection{Hybrid CFD-DSMC}
\label{subsec:hybridmethod}

\begin{figure}[ht]
    \centering
    \includegraphics[width=\textwidth]{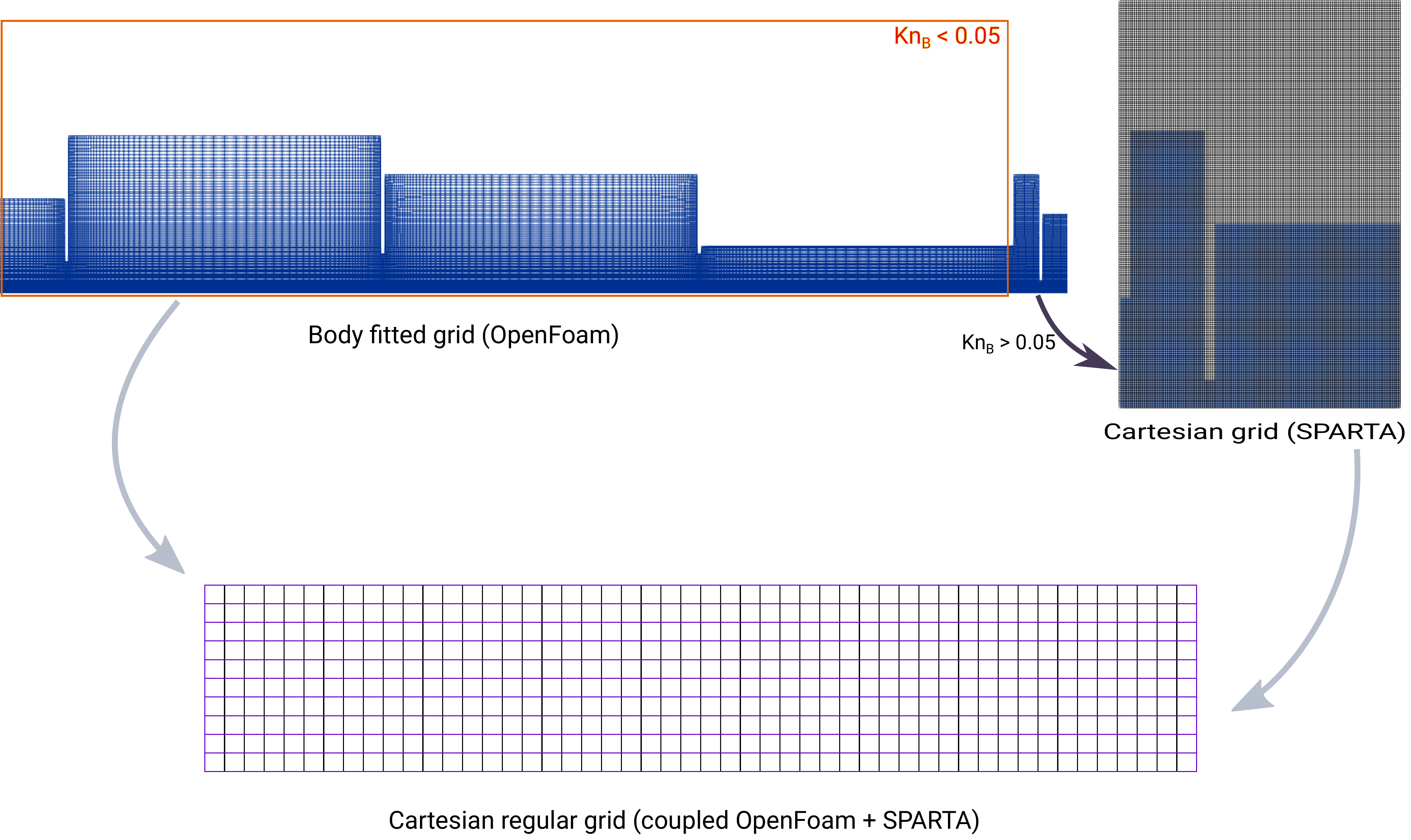}
    \caption{Schematic diagram showing the hybrid CFD/DSMC coupling.}
    \label{fig:cfd-dsmc_coupling}
\end{figure}

The DSMC method had demonstrated the capability of resolving rarefied flows. However, for low Knudsen number flows this approach is computationally very expensive due to a drastic increase in collisions between the molecules. Furthermore, a large number of simulations must be carried out to filter out the statistical noise, which is observed particularly in the low-speed regions of the flow. As described
earlier, the Navier-Stokes solution had always been a better choice in this regime both in terms of
accuracy and efficiency. For experiments with intermediate pressures at the ALS inlet, i.e.,
$50~\text{Pa}\leq~p_{in}\leq180$~Pa, it was observed that the flow through the ALS has a variable
Knudsen number regime, i.e., it changes from continuum to transition and free-molecular-flow
regime. Therefore, we setup a coupled CFD-DSMC approach for resolving such flows. The flow is
initially simulated using CFD (Navier-Stokes) and a continuum breakdown criterion is evaluated.
Based on this criterion, the former computational domain is split into CFD and DSMC regions using an
interface. At this interface, the CFD solution data (flow variables) are interpolated and this
interpolated data are used to generate the required inflow molecular flux per unit
time~\cite{Bird1994} for the DSMC domain to carry out the DSMC simulation in the rarefied region.
The DSMC solution is sampled to extract macroscopic information, e.g., velocity, pressure and
temperature, of the flow and the statistical noise is filtered out. The steady-state solution of the
flow from both CFD, with a body-fitted grid, and DSMC, with a Cartesian grid, in their specific
regions are then interpolated together on a regular Cartesian grid, see
\autoref{fig:cfd-dsmc_coupling}, to have a smoothed contiguous multi-scale flow field. This one-way coupled hybrid method was validated on a gas-dynamic nozzle and the results
showed higher accuracy and computational efficiency than the pure DSMC
method~\cite{peravali2024accuracy}.

To estimate the continuum breakdown criterion in this approach, two different definitions of the Knudsen number are utilized: (a) Global Knudsen number $\Kn$; (b) Local or Boyd's gradient length Knudsen number $\Kn_{GLL,Q}$:
\begin{equation}\label{eq:knud}
    \Kn = \frac{\lambda}{L} \; ; \qquad  \Kn_{GLL,Q} = \frac{\lambda |\nabla Q|}{Q}.
\end{equation}

Here, $\lambda$ represents the mean free path of the gas, $L$ is the characteristic length scale, $Q$ represents a macroscopic flow property such as the density $\rho$, the velocity $\textbf{v}$ or the temperature $T$. The breakdown parameter $\Kn_B$ is estimated based on the maximum of the global and local Knudsen numbers over the computational domain and is compared to a threshold limit of $0.05$. If the value of $\Kn_B$ gets larger than this limit, the region is dedicated to DSMC.
\begin{equation}\label{eq:Kn_break}
        \Kn_B = \max(\Kn, \Kn_{GLL,\rho}, \Kn_{GLL, T}, \Kn_{GLL,|\textbf{v}|}).
    \end{equation}

\subsection{Particle transport}
\label{subsec:particle}

Particle trajectories are calculated using the Langevin approach where the forces on the
nanoparticles are computed as the sum of the drag force $\textbf{F}_\text{drag}$ and the Brownian
motion force $\textbf{F}_b$:
\begin{equation}
    \frac{d }{dt} \left(m_p \, \textbf{u}_p \right) = \textbf{F}_\text{drag} + \textbf{F}_b \; ,
\end{equation}

with the mass of the particle $m_p$, the particle velocity vector $\textbf{u}_p$, and time $t$. In
the following,different models for the drag force are described.

\subsubsection{Stokes-Cunnningham drag model}

The conventional Stokes drag is corrected by the
Cunningham slip coefficient ($C_c$)~\cite{cunningham1910}:

 \begin{equation}\label{eq:stokes}
        \textbf{F}_\text{drag} = \frac{6 \, \pi \, \mu \, r_p \, \Delta \textbf{U}}{C_c}
    \end{equation}

with 

    \begin{equation}\label{eq:cunningham}
        C_c = 1 + \Kn_p \; [A_1 + A_2 \exp (-A_3/\Kn_p)]
    \end{equation}

where $\mu$ is the dynamic gas viscosity, $r_p$ is the radius of the particle and $\Delta \textbf{U}$ is the difference in velocity between the gas and the particle. For calculating $C_c$, the particle Knudsen number $\Kn_p$ is defined as the ratio of the mean free path of the gas to the radius of the particles. The coefficients $A_1 = 1.231$, $A_2=0.4695$ and $A_3=1.1783$ were empirically obtained~\cite{Hutchins1995}. A further correction to this model for high Mach number flows was also provided~\cite{Nsingh_Tom2021}.

\subsubsection{Molecular drag models}\label{sec:molec_drag}

To reduce the dependence on empirical coefficients, analytically derived models were considered for the extremely rarefied regimes in this study. When the size of the nanoparticle is very small compared to the mean free path of the gas and gas atoms/molecules are specularly reflected from the surface of the nanoparticle, the drag force of the nanoparticle is~\cite{Epstein}:
\begin{equation}\label{eq:epstein}
        \textbf{F}_\text{spec} = \frac{4 \, \pi}{3} \, r_p^2 \, N \,  m \, \overline{c} \, \Delta \textbf{U}  \;  . 
    \end{equation}

Alternatively, for diffusively reflected gas molecules, the drag force is:
\begin{equation}\label{eq:epstein2}
        \textbf{F}_\text{diff} = \left ( 1 + \frac{\pi}{8}\right ) \, \frac{4 \, \pi}{3} \, r_p^2 \, N \,  m \, \overline{c} \, \Delta \textbf{U} \; .
    \end{equation}

Here, $N$ is the number density of the gas molecules, $m$ is the molecular mass of the gas and $\overline{c}$ is the average speed of molecules in the gas. These models are well suited for low-speed flows, i.e., low Mach number flows ($Ma < 0.3$).

An analytical expression for both specularly and diffusively reflected atoms/molecules for
intermediate and high-speed flows (high Mach numbers) is\cite{Baines}:
\begin{equation}\label{eq:baines_spec}
        \textbf{F}_\text{spec} = \frac{\pi^{3/2} \, \rho \, r_p^{2} \, \overline{c}^2}{4} \left \{ \left ( S + \frac{1}{2S} \right )\exp (-S^2) + \sqrt{\pi} \left ( S^2 + 1 - \frac{1}{4S^2} \right ) \operatorname{erf} {S} \right \} \; ,
    \end{equation}

\begin{equation}\label{eq:baines_diff}
        \textbf{F}_\text{diff} = \frac{\pi^{3/2} \, \rho \, r_p^{2} \, \overline{c}^2}{4} \left \{ \left ( S + \frac{1}{2S} \right )\exp (-S^2) + \sqrt{\pi} \left ( S^2 + 1 - \frac{1}{4S^2} \right ) \operatorname{erf} {S} + \frac{\pi S}{3}\right \} \; .
    \end{equation}

Here, $\rho$ is the density of the gas and $S=\sqrt{\frac{m}{2 k_B T}} \cdot \Delta \textbf{U}$ denotes the molecular speed ratio, where $k_B$ is the Boltzmann constant and $T$ the temperature of the gas. The total drag force on the particle is assumed to be a combination of a certain fraction ($\alpha$) of diffuse reflections and the remaining fraction ($1-\alpha$) are specular reflections:
\begin{equation}\label{eq:f_total}
    \textbf{F}_\text{drag} =  (1-\alpha) \; \textbf{F}_\text{spec} + \alpha \; \textbf{F}_\text{diff}
\end{equation}

It is typically assumed that $\alpha=0.9$~\cite{Epstein, millikan1923_1, millikan1923_2,
   ALLEN1982537}, which we also used in the current work.

\subsubsection{Relaxation of Epstein drag}\label{sec:drag_rare_mc}

For particles traversing across low-speed transition or molecular flow regimes (i.e., DSMC regions with $\Kn > 0.05$), we observed by comparison with experimental data that the above-mentioned models overpredict the drag force in this regime due to the overestimation of impinging gas molecules that transfer momentum to the nanoparticle.  Therefore, a relaxation of the drag force is necessary to accurately track particles in the flow by estimating the actual fraction of colliding molecules when particles move through a sub-cell of the simulation domain. For this purpose, a sub-cell of the flow field, in which a certain number of gas molecules exist, is considered. The gas velocity distribution functions in this sub-cell are assumed to follow the Maxwell-Boltzmann distribution. Like in DSMC, certain numbers of simulation molecules are created where each particle represents real molecules in the system that roughly have the same position and velocity. From the macroscopic flow data, such as pressure, flow velocity and temperature, velocities are assigned to the simulation molecules in the sub-cell. The relative velocity of the randomly chosen simulation molecule with respect to the nanoparticle is estimated by: 
\begin{equation}
    \textbf{u}_{r,i} = (\textbf{u}_i + \textbf{U}) - \textbf{u}_{p}
\end{equation}

where $\textbf{u}_i$ is the thermal velocity of the randomly chosen simulation molecule from the Maxwell-Boltzmann distribution, $\textbf{U}$ is the bulk velocity of the gas flow obtained from DSMC and  $\textbf{u}_{p}$ is the velocity of the nanoparticle.

 The collision between the nanoparticle and the impinging gas molecules that have a relative velocity less than the most probable speed of the gas molecules $\beta=\sqrt{2 k_B T/m}$ in the low-speed high-Knudsen number regime is assumed stochastic. Thus, a gas molecule collides with the nanoparticle, if
 \begin{equation}\label{eq:coll_frac}
     1 - \exp \left (- \frac{|\textbf{u}_{r,i}| }{\beta} \right) > R_f  \; .
 \end{equation}

Here, $R_f$ is a randomly generated number from $(0,1]$ with a uniform distribution and  Eq.~\eqref{eq:coll_frac} filters certain impinging molecules using Monte-Carlo acception-rejection sampling. The fraction of colliding molecules $P_\text{coll}$ is determined per time step and the total drag force $\textbf{F}_\text{drag}$ from Eq.~\eqref{eq:f_total} (obtained from Eqs.~\eqref{eq:epstein}~and~\eqref{eq:epstein2}) is relaxed accordingly:
\begin{equation}\label{eq:drag_relaxed}
    \textbf{F}_\text{drag,~relaxed} = P_\text{coll} \cdot \textbf{F}_\text{drag} \; .
\end{equation}


\subsubsection{Brownian motion}

The drag force estimated above is the force obtained by averaging single collisions undergone by the particle per unit time, i.e., it is the mean force acting on the particle. However, the particle trajectory is also influenced by the Brownian motion due to the nanometer size range of the particle. The Brownian motion force is defined based on a Gaussian white noise random process having a spectral intensity $S_0$ as:
\begin{equation}\label{eq:brown1}
    \textbf{F}_b = m_p \; \textbf{G} \; \sqrt{\frac{\pi \, S_0}{\Delta t}} \; .
\end{equation}

Here, $\textbf{G}$ is a vector of independent Gaussian random numbers with zero mean and unit variance and $\Delta t$ is the time step. For the drag force modeled with the Stokes-Cunningham relation, the spectral intensity is defined as~\cite{Li_ahmedi}:
\begin{equation}
    S_0 = \frac{27 \, \mu \, k_B T}{4 \, \pi^2 \, r_p^5 \, \rho_p^2 \, C_c} \; ,
\end{equation}

where $\mu$ is the dynamic viscosity and $\rho_p$ the density of the particle. For the molecular drag force model the spectral intensity is calculated as~\cite{roth2020microscopic}:
\begin{equation}\label{eq:brown2}
    S_0 = \left ( \frac{16}{3} + \frac{2\pi}{3} \sqrt{\frac{T_p}{T}}\right )\frac{\overline{c}}{2} \, p \,  \frac{m}{m^2_p} \,  r_p^2 \; .
\end{equation}

Here, $T_p$ is the temperature of the particle and $p$ is the pressure of the gas.

\section{Results and discussion}
\label{sec:results}

\begin{figure}[!hbt]
    \hspace*{-16mm}\includegraphics[width=210mm]{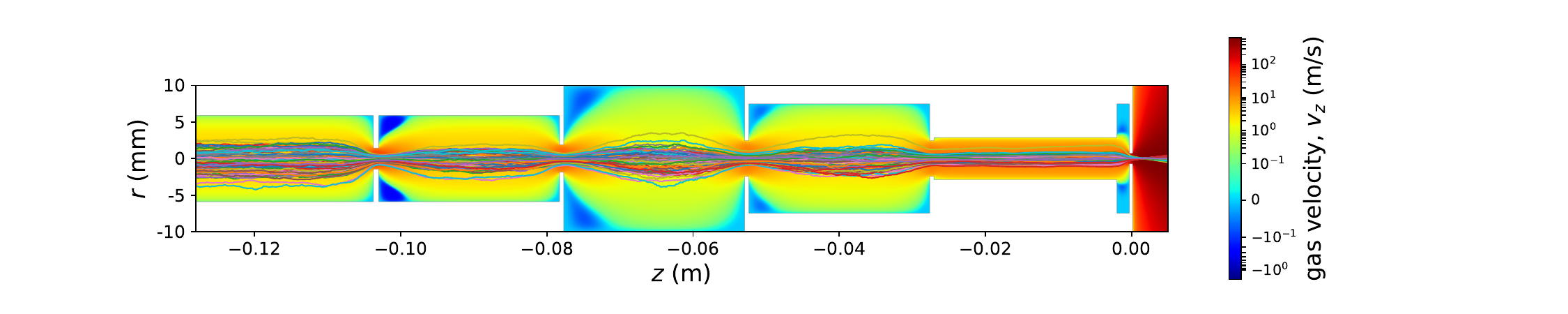}
    \caption{Simulated trajectories (colored lines) of the 25~nm PS through the aerodynamic-lens stack drawn on top of the gas-flow field ($p_{in}=150$~Pa) visualized by the axial velocity in a logarithmic color scale.}
    \label{fig:traj}
\end{figure}

The numerical methodologies described in~\autoref{sec:method} are utilized to simulate the particle-beam evolution at different conditions presented in~\autoref{tab:flowcases}. \autoref{fig:traj} exemplarily shows the flow field and the corresponding nanoparticle trajectories throughout the computational domain for the 25~nm polystyrene spheres (PS)  at an inlet pressure of $p_{in} = 150$~Pa. Here, $r$ represents the radial and $z$ the axial coordinate of the flow domain. The flow field predicted by the hybrid CFD-DSMC method is depicted by the axial velocity $v_z$ representing the main flow direction and the particle trajectories are calculated by molecular drag force models (\autoref{sec:molec_drag}).

\begin{figure}[!hbt]
	\centering
    \begin{subfigure}{0.49\textwidth}
      \centering
		\includegraphics[width=\textwidth]{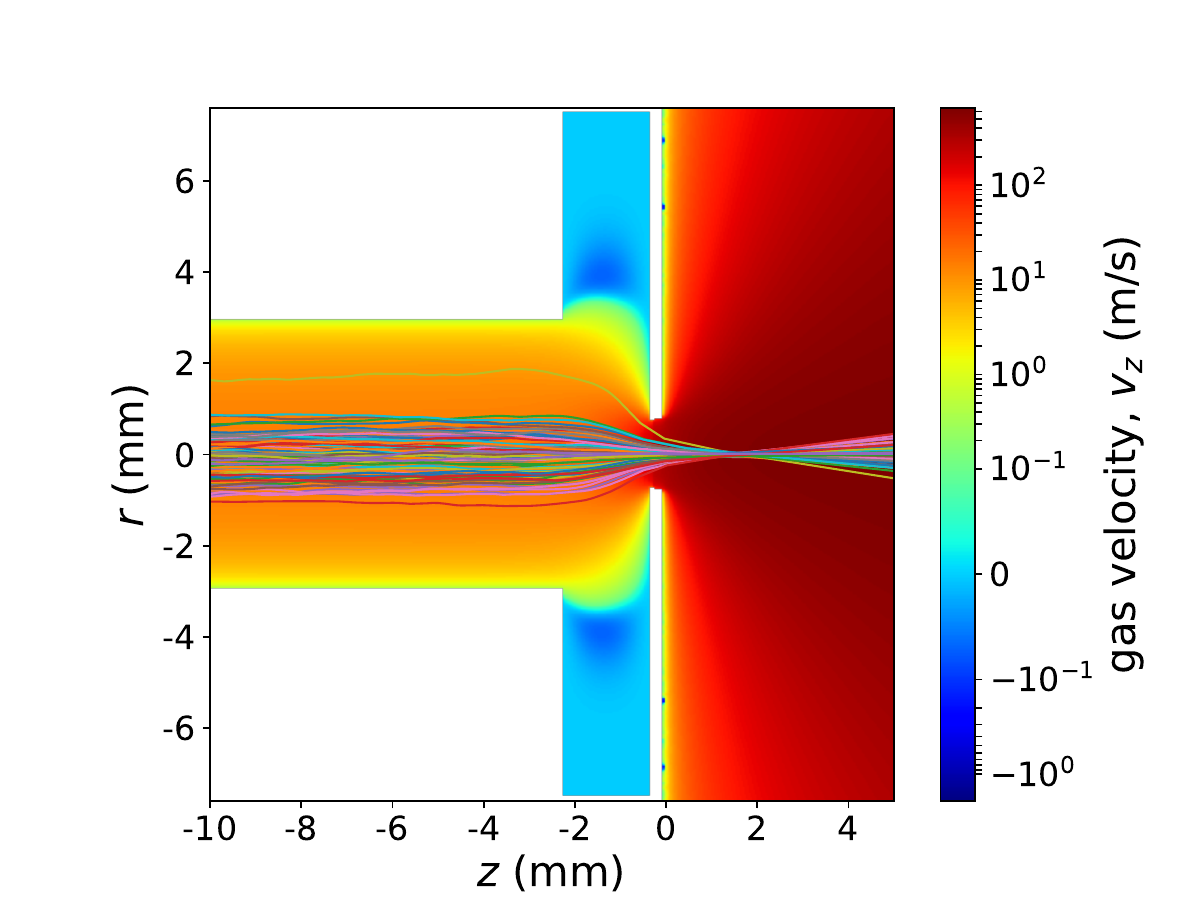}
        \subcaption{\label{subfig:traj_zoom_a}}
     \end{subfigure}
     \begin{subfigure}{0.49\textwidth}
      \centering
        \includegraphics[width=\textwidth]{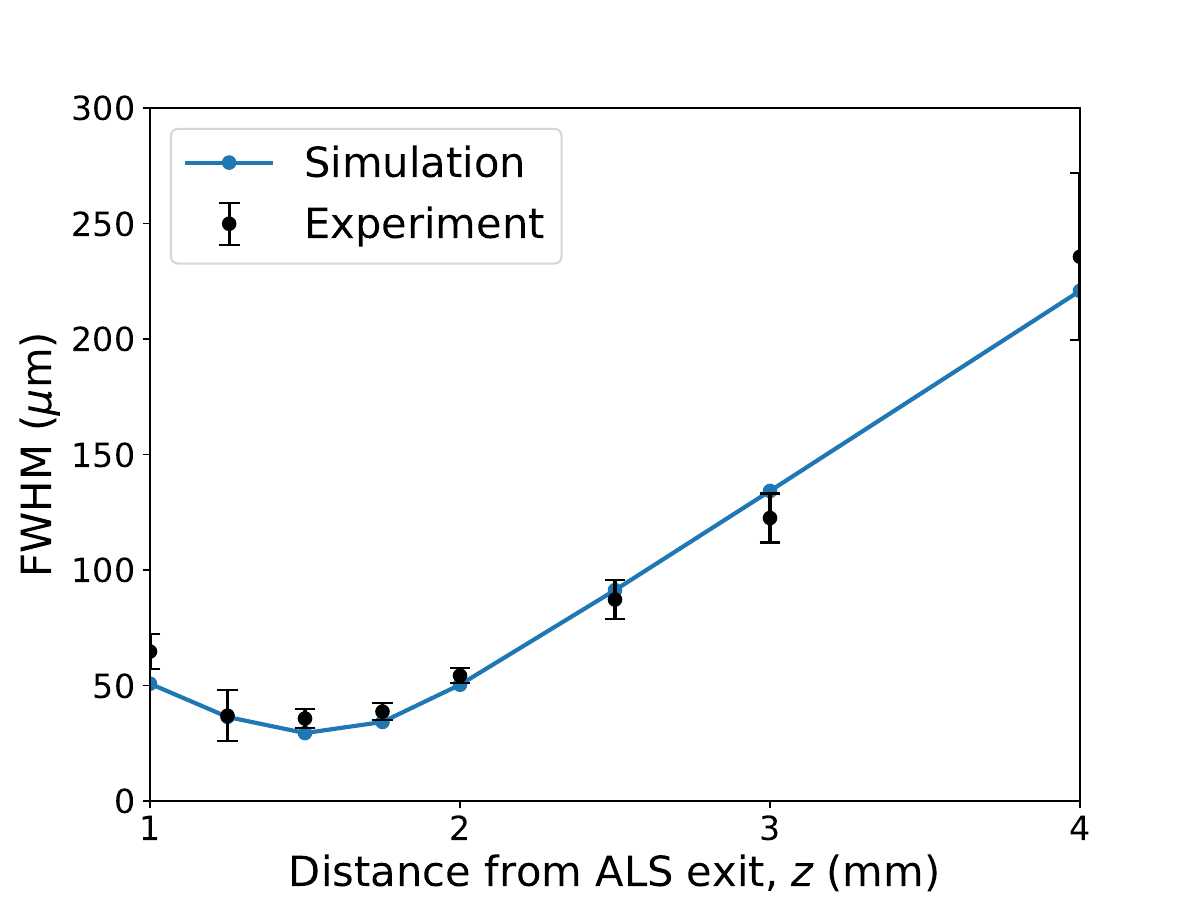}
        \subcaption{\label{subfig:traj_zoom_b}}
     \end{subfigure}
	\caption{(a) Zoom-in view of the ALS-exit into the high-vacuum chamber of \autoref{fig:traj},
      including simulated particle trajectories from the ALS exit into vacuum;
    (b) Particle-beam-size evolution (FWHM) of $25$~nm PS
       at an inlet pressure of  $p_{in}= 150$~Pa.}
	\label{fig:traj_zoom}
\end{figure}

 The exit of the ALS is at $z = 0$. The flow domain defined by $z>0$ represents the vacuum chamber, where the gas flow from the ALS expands at supersonic speeds. \autoref{subfig:traj_zoom_a} is the zoomed-in view of~\autoref{fig:traj}, which shows the simulated particle trajectories focusing (converge to a minimum beam width) and de-focusing inside the vacuum chamber. In the vacuum chamber, the particle beam widths are measured at different positions starting at $z = 1$~mm and onwards. The particle beam evolving from the exit of the aerodynamic lens has a Gaussian-like distribution~\cite{worbs_phd2022}. Therefore, the width of the particle beam is designated by the full-width at half-maximum (FWHM). The widths of the simulated particle beam at the corresponding experimental positions are compared with the experimental data in~\autoref{subfig:traj_zoom_b}. In the current case, the beam profiles obtained by simulation show good agreement with the experimental data.
 
 For every experimental case in~\autoref{tab:flowcases}, $10^4$
 particles were simulated with an initial radial velocity of $v_r = 0$
 and an axial velocity following a normal distribution with a zero mean
 and a standard deviation of 10~m/s.
 The particles are positioned at the ALS inlet with a Gaussian distribution
 centered around $r = 0$ and FWHM of 0.0023~m. 
 For particle numbers above 1000, the simulated beam profiles
 do not change significantly. Thus, with ten times more 
 particles, it is ensured that the statistics are fully converged.
 

However, the level of accuracy of the simulation shown above highly depends on choosing the right modeling approach, which also varies for different experimental conditions such as inlet pressure or particle size. In the following subsections, the numerical approaches and models for the drag force mentioned in~\autoref{sec:method} are evaluated for different flow conditions by comparing the experimental particle beam evolution in the vacuum chamber with the simulated particle beam profile.  

\subsection{Multi-scale regime \label{subsect:multiscale}}

For the test cases having variable Knudsen number regimes, the flow field is simulated using the hybrid CFD/DSMC approach described in~\autoref{subsec:hybridmethod}. 

The particle trajectories are subsequently predicted and evaluated like above for 69~nm and 42~nm PS. Since the test cases have very high particle Knudsen numbers (see~\autoref{tab:flowcases}), the corresponding molecular drag force models are chosen based on the Mach number ($Ma$) of the flow. For $Ma <0.3$, the \citet{Epstein} model (Eqs.~\eqref{eq:epstein}~and~\eqref{eq:epstein2}) is used and for $Ma > 0.3$ the drag model switches to \citet{Baines} (Eqs.~\eqref{eq:baines_spec}~and~\eqref{eq:baines_diff}). However, it is observed that there are no significant deviations between the results achieved with these molecular drag models and the Stokes-Cunningham model (along with the correction to high Mach number flows~\cite{Nsingh_Tom2021}) .

\begin{figure}[!hbt]
	\centering
    \begin{subfigure}{0.49\textwidth}
      \centering
		\includegraphics[width=\textwidth]{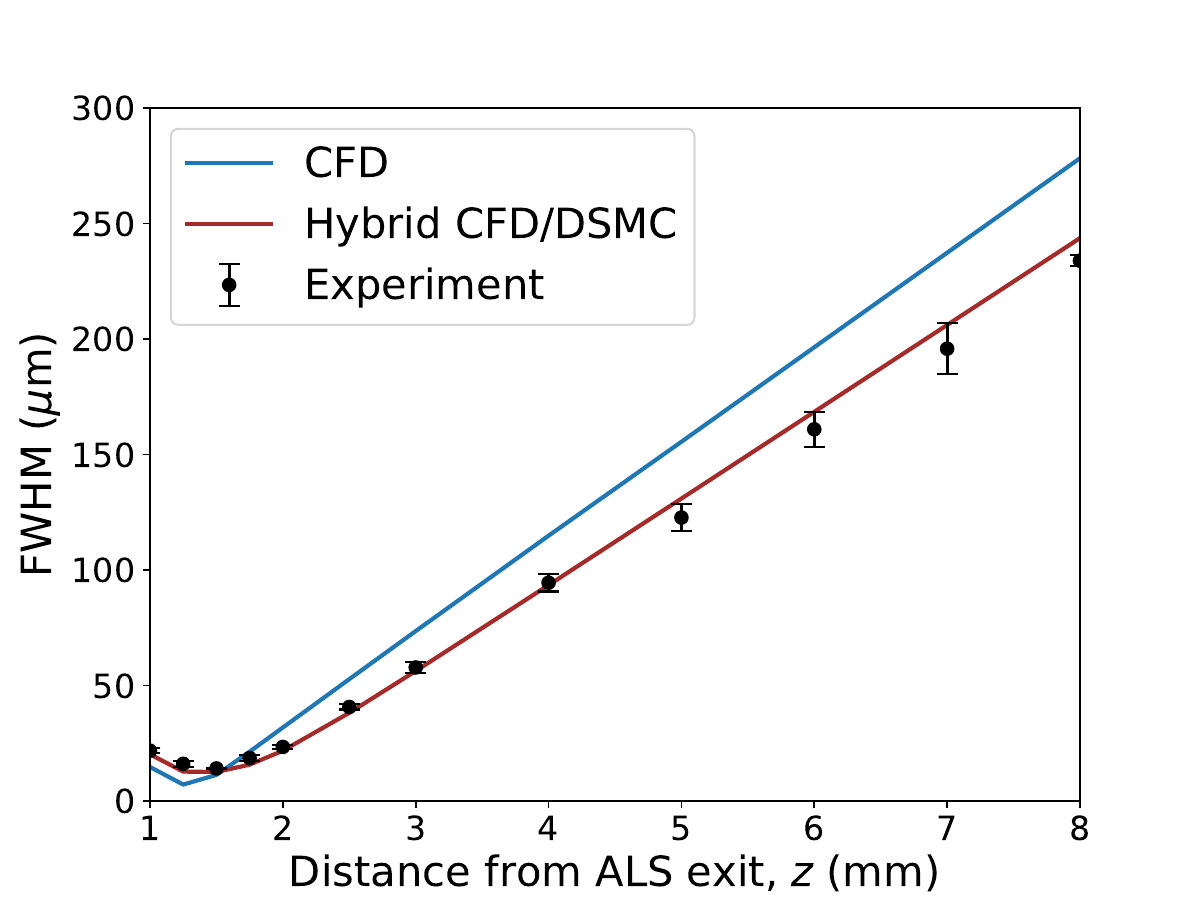}
        \subcaption{\label{subfig:69_a} $180$~Pa}
     \end{subfigure}
     \begin{subfigure}{0.49\textwidth}
      \centering
        \includegraphics[width=\textwidth]{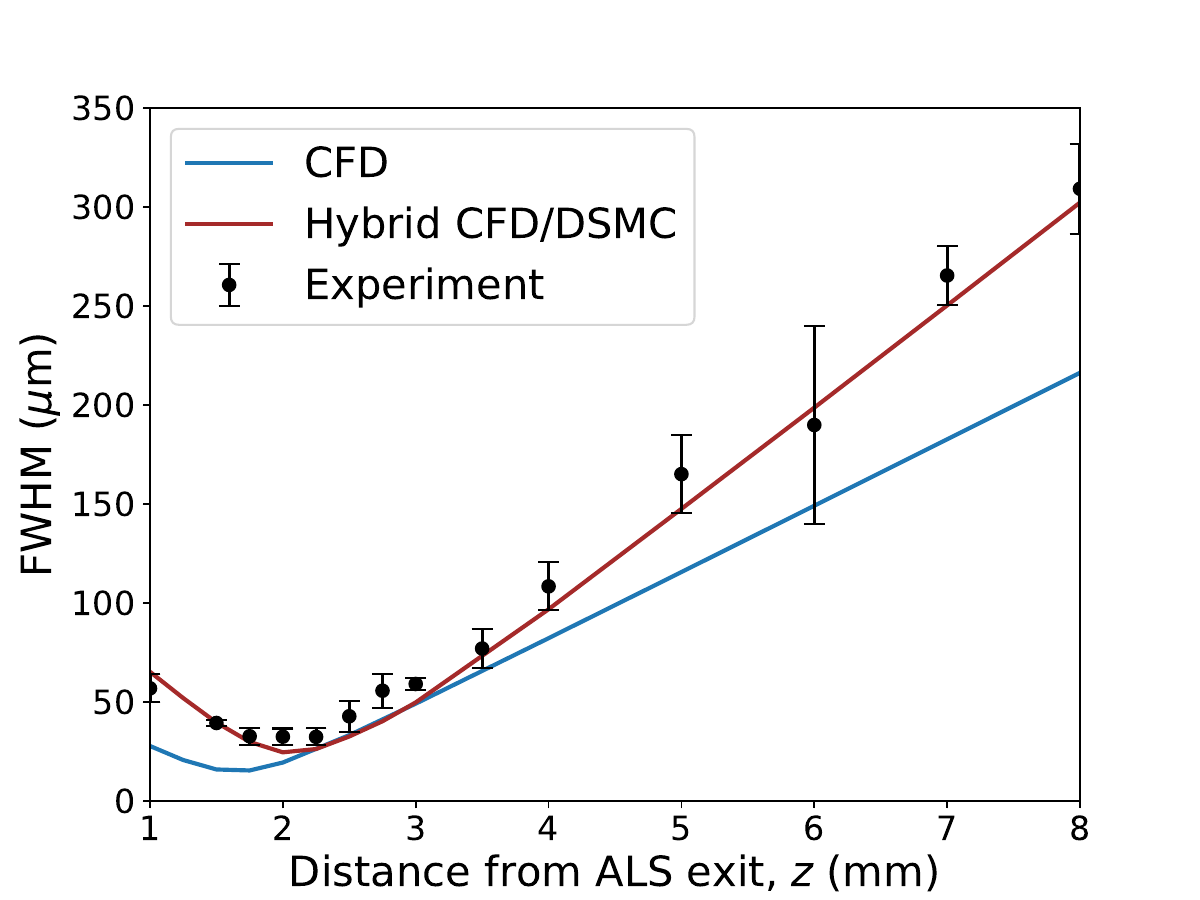}
		  \subcaption{\label{subfig:69_b} $55$~Pa}
     \end{subfigure}
	\caption{Particle-beam-size evolution (FWHM) of $69$~nm PS at two different inlet pressures.}
	\label{fig:hybresults_69}
\end{figure}

\begin{figure}[!hbt]
	\centering
	\begin{subfigure}{0.49\textwidth}
      \centering
		\includegraphics[width=\textwidth]{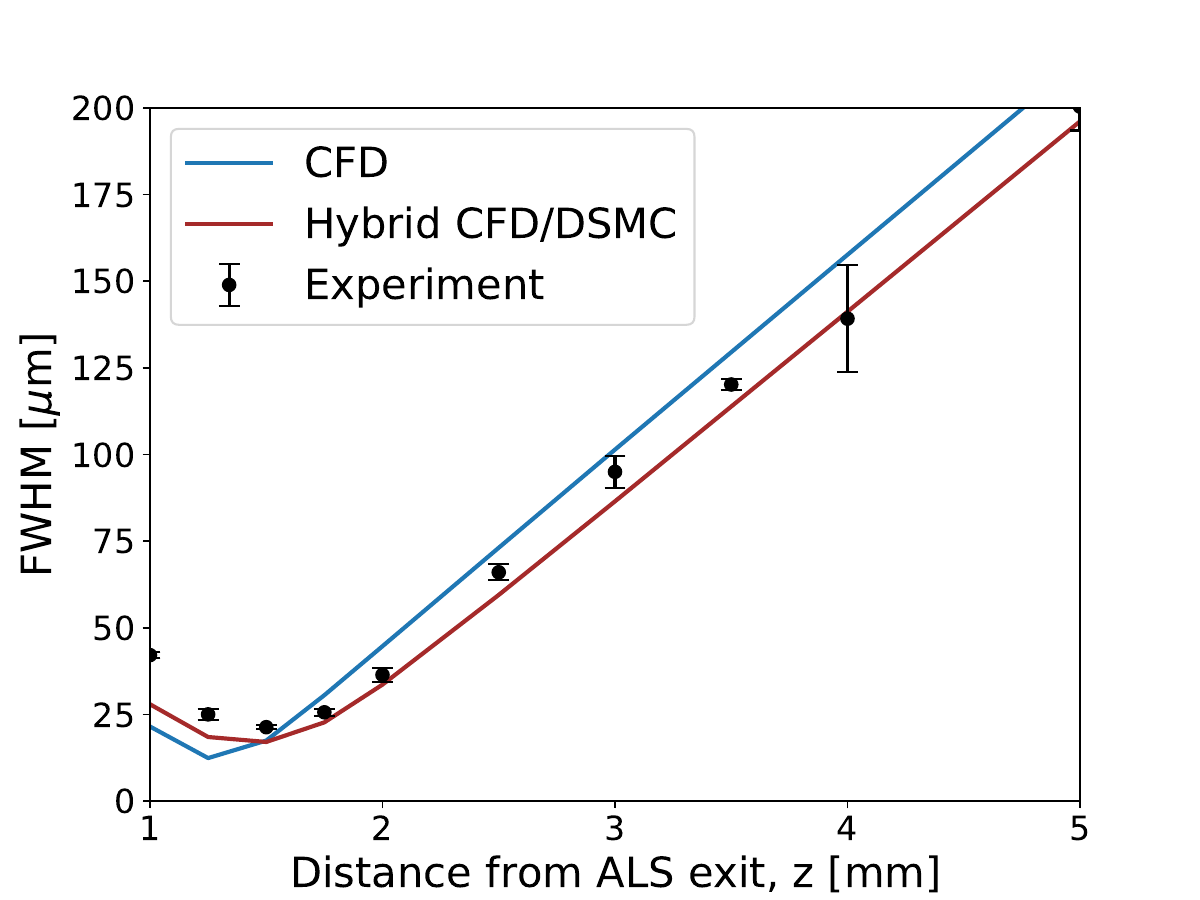}
        \subcaption{\label{subfig:42_a} $180$~Pa}
     \end{subfigure}
     \begin{subfigure}{0.49\textwidth}
      \centering
        \includegraphics[width=\textwidth]{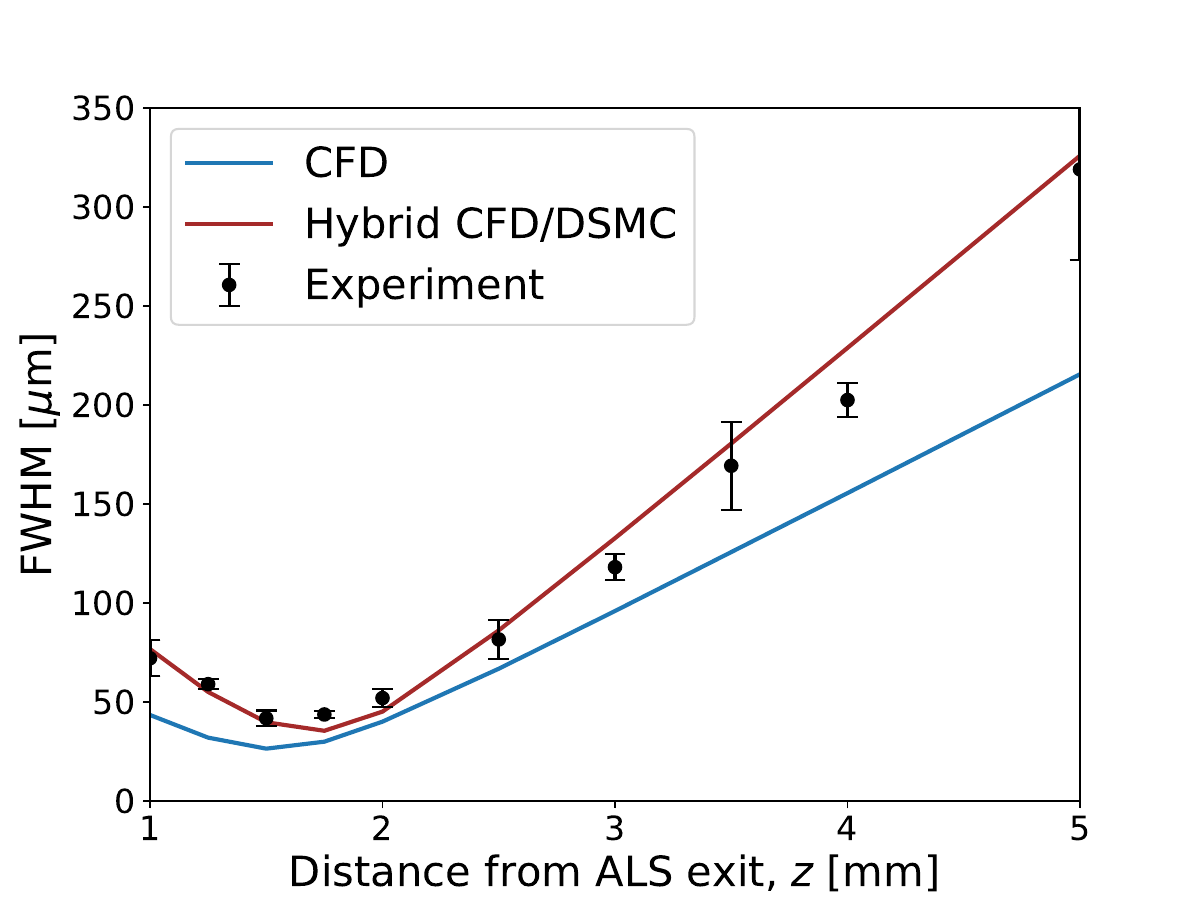}
     \subcaption{\label{subfig:42_b} $55$~Pa}
     \end{subfigure}
	\caption{Particle-beam-size evolution (FWHM) of $42$~nm PS at two different inlet pressures.}
	\label{fig:hybresults_42}
\end{figure}

\begin{figure}[!hbt]
   \centering
   \begin{subfigure}{0.49\textwidth}
     \centering
        \includegraphics[width=\textwidth]{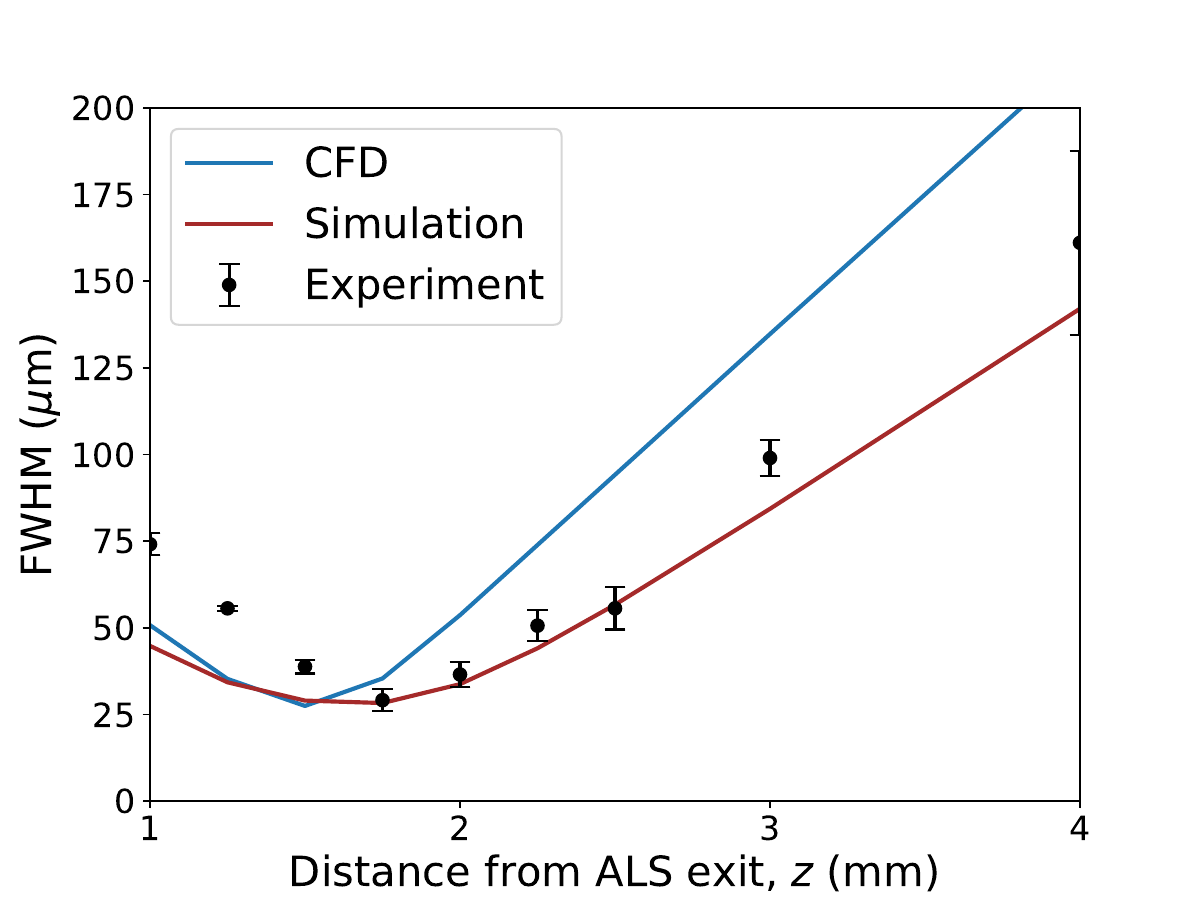}
        \subcaption{\label{subfig:25nm_a} $200$~Pa}
   \end{subfigure}
   \begin{subfigure}{0.49\textwidth}
     \centering
        \includegraphics[width=\textwidth]{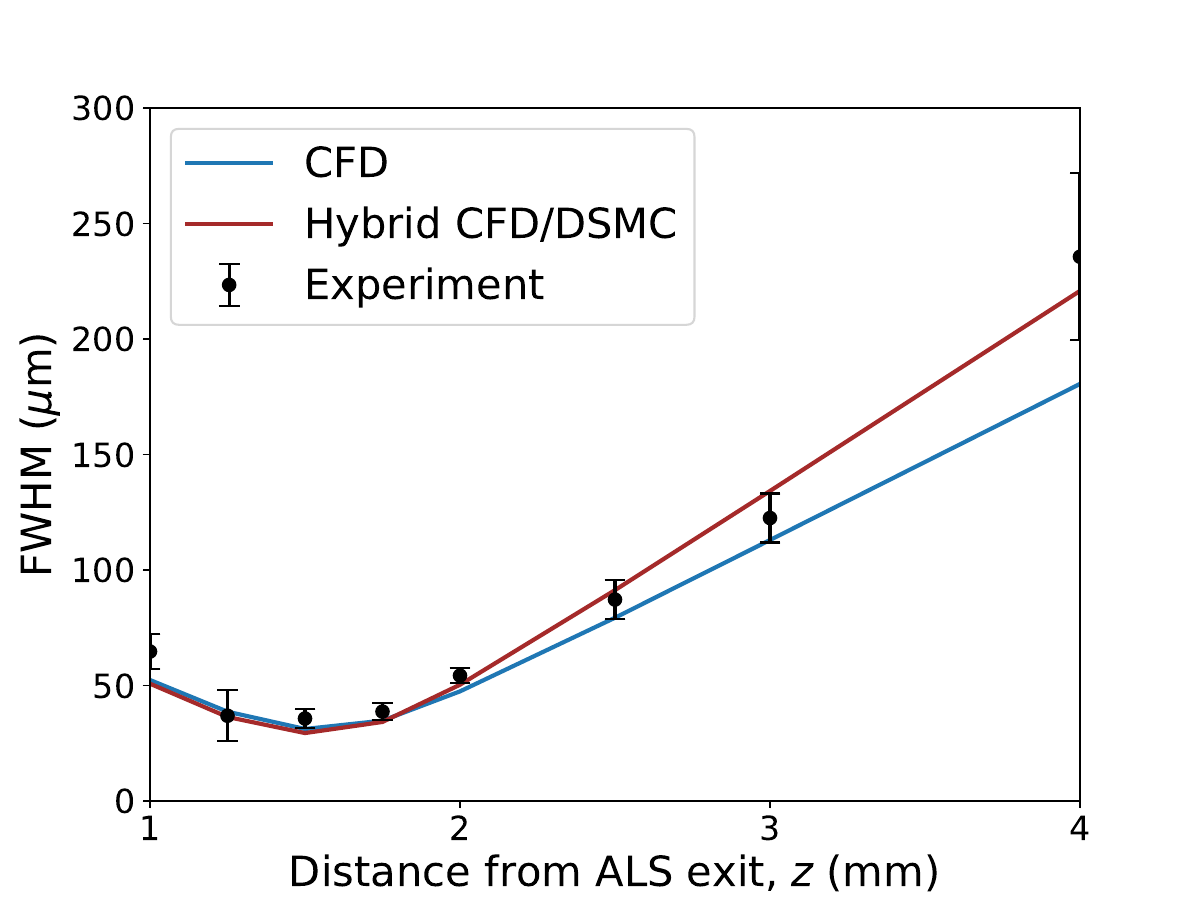}
        \subcaption{\label{subfig:25nm_b}  $150$~Pa }
   \end{subfigure}
   \begin{subfigure}{0.49\textwidth}
     \centering
        \includegraphics[width=\textwidth]{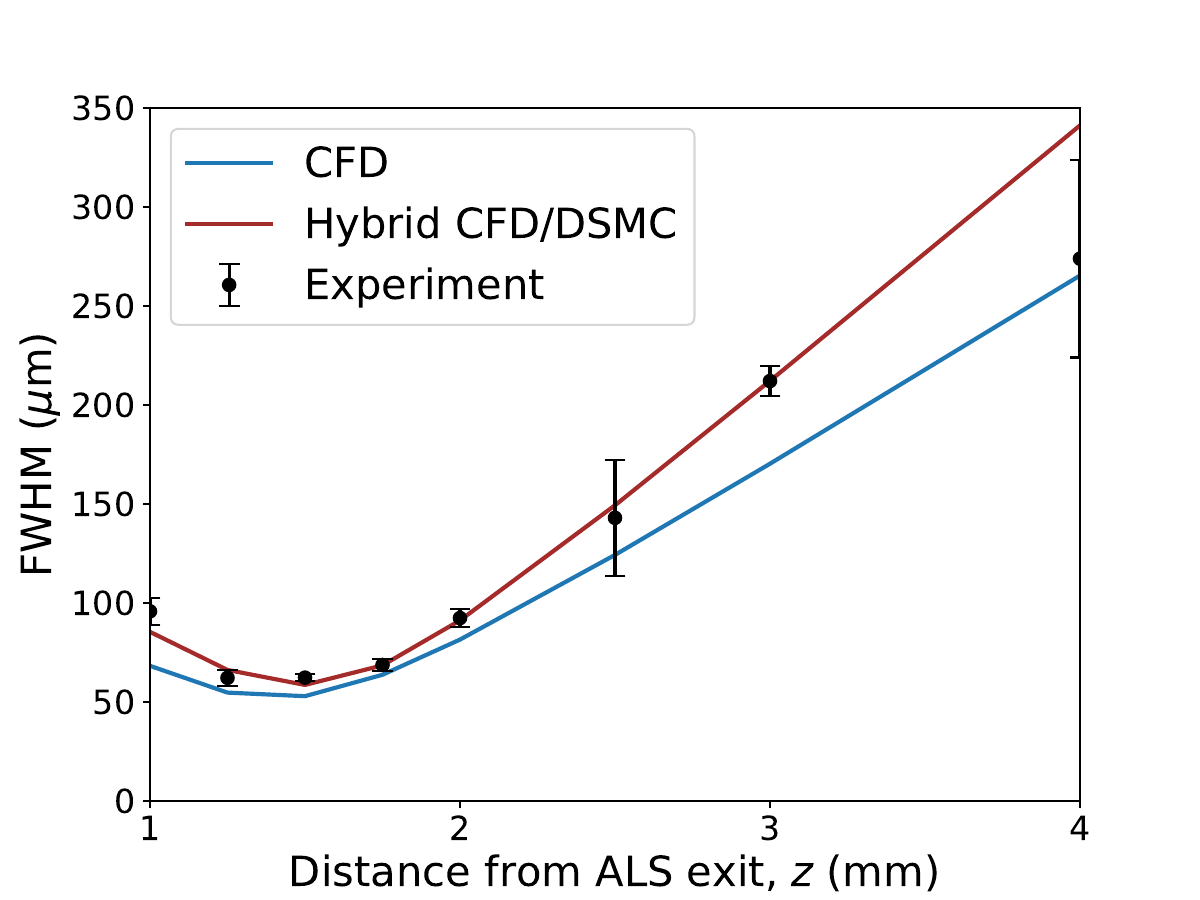}
        \subcaption{\label{subfig:25nm_c} $50$~Pa}
   \end{subfigure}
  \caption{\label{fig:hybresults_25}  Particle-beam-size evolution (FWHM) of
             $25$~nm PS at three different inlet pressures. }
\end{figure}

\autoref{fig:hybresults_69}, \autoref{fig:hybresults_42}, and \autoref{fig:hybresults_25} show the particle beam widths at different positions behind the ALS exit for particle sizes of 69~nm, 42~nm and 25~nm, respectively. The predicted data are given for different inlet pressures shown in~\autoref{tab:flowcases}. 
The results predicted by the hybrid CFD/DSMC method show very good agreement with the experimental data for all particle sizes (i.e., focusing-defocusing behavior and focus position) compared with the pure CFD.  The particle beam widths computed based on the pure CFD code deviate significantly from the experiment as the inlet pressure reduces. Additionally, Appendix~\ref{app:additinal_cases} shows a case with 25~nm gold nanoparticles (AuNP) where the hybrid DSMC-CFD methodology could predict the experimental trend
quite well, too.

\subsection{Highly rarefied regime \label{subsect:rarefied}}

For the test cases with an inlet pressure of $p_{in}=20~\text{Pa}$ mentioned in~\autoref{tab:flowcases}, the maximum global and local Knudsen numbers are evaluated to be greater than $0.1$. Therefore, for these test cases it is ideal to use the pure DSMC approach. 
Once a smooth sampled flow field is established using DSMC, the particle trajectory calculations are carried out for 69~nm and 42~nm PS. Like in the cases described in the previous section, the corresponding molecular drag force models are chosen based on the Mach number of the flow (Eqs.~\eqref{eq:epstein} / \eqref{eq:epstein2}) or Eqs.~\eqref{eq:baines_spec} / \eqref{eq:baines_diff}). Furthermore, the relaxed drag force model based on Monte-Carlo sampling (Eq.~\eqref{eq:drag_relaxed}) described in~\autoref{sec:drag_rare_mc} has also been used in place of the \citet{Epstein} model.
  
\begin{figure}[!hbt]
	\centering
    \begin{subfigure}{0.49\textwidth}
     \centering
		\includegraphics[width=\textwidth]{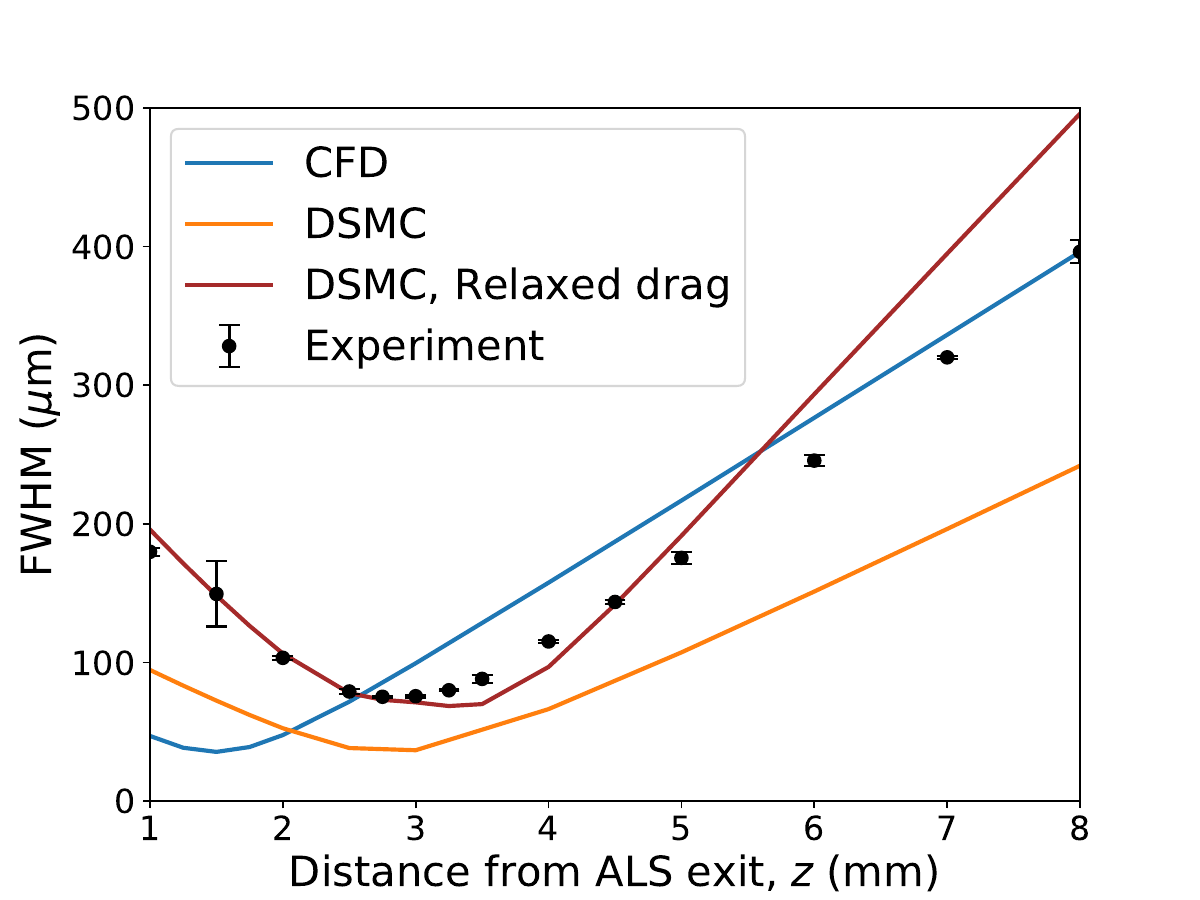}
        \subcaption{\label{subfig:20Pa_69}   $69$~nm}
   \end{subfigure}
   \begin{subfigure}{0.49\textwidth}
     \centering
     \includegraphics[width=\textwidth]{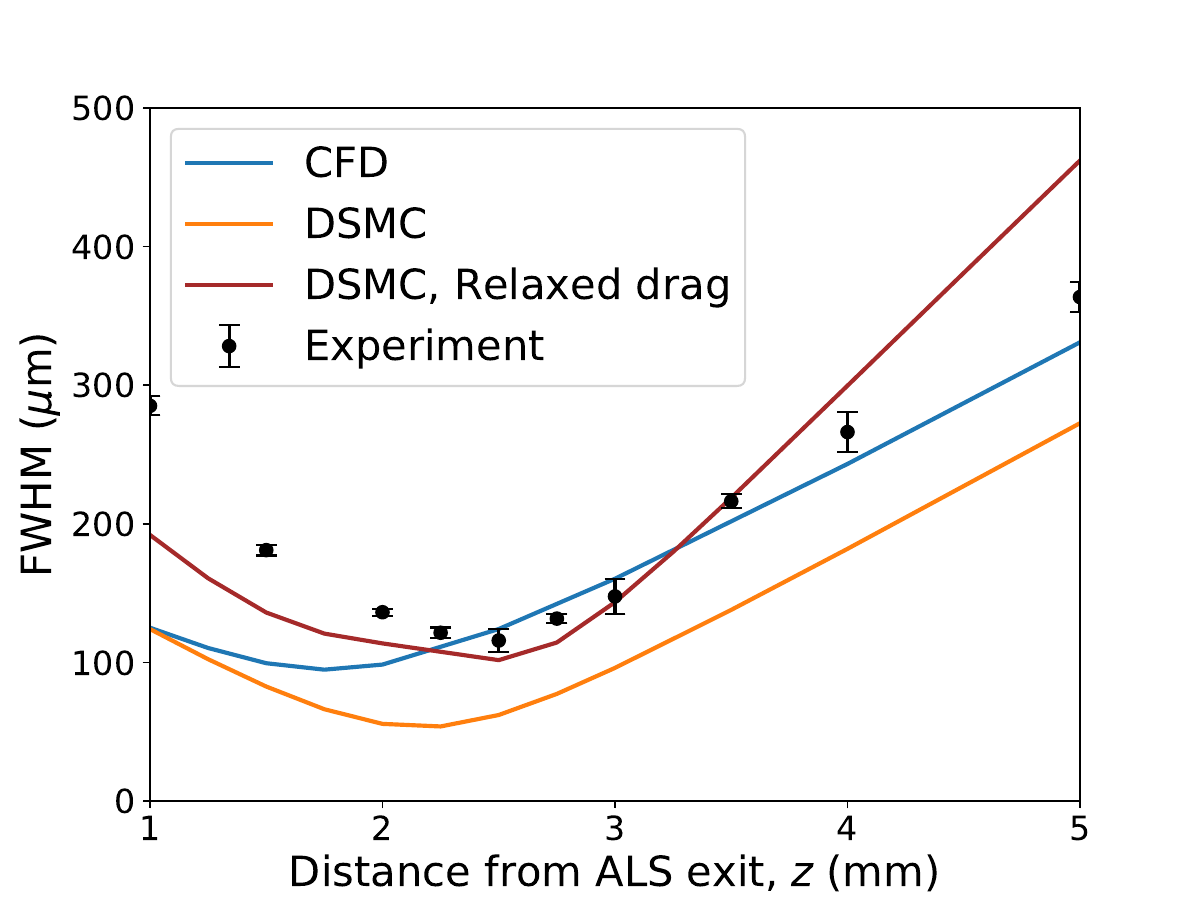}
	\subcaption{\label{subfig:20Pa_42}  $42$~nm}
   \end{subfigure}	
   \caption{Particle-beam-size evolution (FWHM) of PS
            at an inlet pressure of 20~Pa for two different particle sizes.}
	\label{fig:dsmc_drag}
\end{figure}

 \autoref{subfig:20Pa_69}~and~\autoref{subfig:20Pa_42} show
 the particle beam widths at different distances from the
 ALS exit for particle sizes of 69~nm and 42~nm, respectively.
 In addition to the pure DSMC method, for comparison purposes 
 the underlying flow fields are also simulated using pure CFD. 

The results predicted by CFD are not in good agreement with the experimental data for both particle sizes. Here, both the focus position and the width of the particle beam are underpredicted. In the case of DSMC, the particle beam evolution shows a similar trend as the experimental data, where the position of the focus is predicted
in reasonable agreement with the measurements. However, the beam widths are underpredicted due to overestimation of the drag force in this regime. Therefore, the molecular drag is relaxed according to Equations~\eqref{eq:coll_frac}~and~\eqref{eq:drag_relaxed}. Obviously, this corrected drag force yields a much better agreement with the experimental data for both particle sizes.

\clearpage

\section{Conclusions}
\label{sec:conclusions}

We proposed and implemented an enhanced and accurate numerical methodology for the simulation of nanoparticle injection through aerodynamic lens systems. Our approach handles both the carrier gas flow through the system and the particle trajectories. For the former, a hybrid molecular-continuum simulation method was set up, which accounts for a wide range of Knudsen numbers in the flow fields of such lens systems, ranging from high-density gas to a highly rarefied flow during the expansion into the vacuum chamber.

Coupling CFD and DSMC allowed for limiting the use of the much more CPU-time intensive molecular model only in those regions, which can not be accurately predicted by the continuum mechanics approach. For the prediction of the particle trajectories, drag force models from the literature were evaluated including molecular drag models. For particles traversing through transitional regimes at the boundary between continuum and molecular flow, an additional correction factor was derived, taking into account the probability that a fraction of the molecules does not collide with a particle in a sub-cell. 

The entire methodology was applied to nine different experimental configuration, three particle sizes and three inlet pressures, spanning a wide parameter space. In the multi-scale regime, the hybrid DSMC/CFD approach proves to be superior to the pure CFD method. No significant deviations between the results achieved with the molecular drag models and the Stokes-Cunningham model were observed. For the highly rarefied cases, the combination of the DSMC approach with the newly proposed relaxation of the drag force led to good agreement with the experimental data~\citep{worbs_phd2022}, which was not the case for the classical models. However, this model requires validation against different gas flow conditions, e.g., multi-species gas, and over a wide range of temperatures, 4 to 300~K, and particle sizes, 1 to 25~nm. 

Future experiments are planned for improving the characterization of the relaxed drag force as well as for generating training data for the development of machine-learning models aimed at improving the semi-empirical drag models across a large range of flow conditions. 
Similarly, efforts will focus on improving heat transfer models that describe particle-gas interactions under varying collision dynamics.

\section{Acknowledgements}

This work was supported by the Helmholtz Data Science Graduate School for the Structure of Matter
(DASHH, HIDSS-0002), by the Helmut-Schmidt University, University of the Armed Forces Hamburg, by
Deutsches Elektronen-Synchrotron DESY, a member of the Helmholtz Association (HGF), and by the
Cluster of Excellence ``Advanced Imaging of Matter'' (AIM, EXC 2056, ID 390715994) of the Deutsche
Forschungsgemeinschaft (DFG).

We acknowledge dtec.bw – Digitalization and Technology Research Center of the Bundeswehr (project
hpc.bw) for provision of computational resources on HSUper, dtec.bw is funded by the European Union
-- NextGenerationEU. We also acknowledge the Maxwell computational resources operated at Deutsches
Elektronen-Synchrotron DESY.

\appendix
\section{Additional case -- Focusing 25~nm gold nanometer particles}
\label{app:additinal_cases}

\begin{figure}[!hbt]
    \centering
    \includegraphics[width=0.5\linewidth]{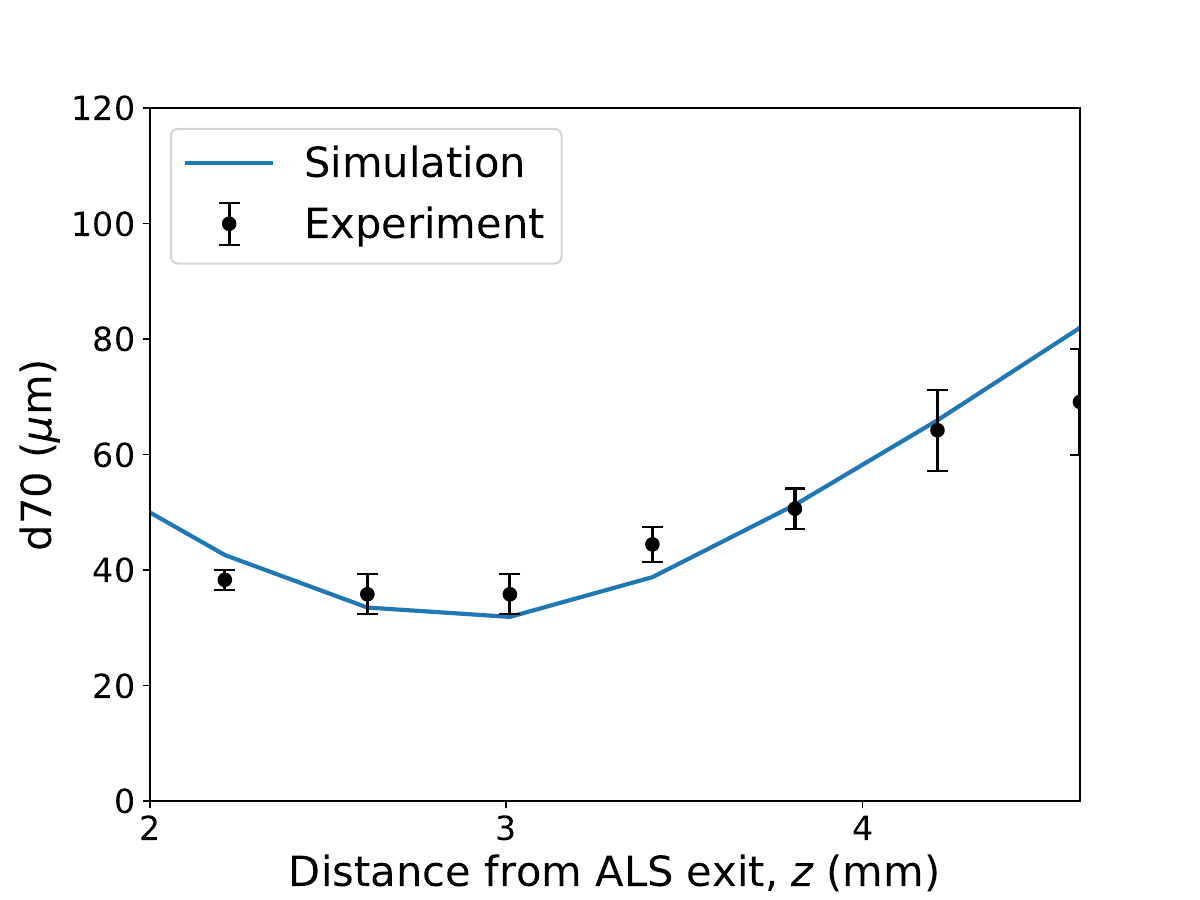}
    \caption{Particle-beam-size evolution of 25~nm AuNP at inlet pressure of 180~Pa.}
    \label{fig:additionalcases}
\end{figure}

In this section, an additional test case is presented. \autoref{fig:additionalcases} shows the particle beam width at different positions after the ALS exit for gold spheres of 25~nm at an inlet pressure of 180~Pa. For this gold-sphere case, a slightly different ALS geometry is used~\cite{Worbs:te5083}. It has to be noted that for this setup, the beam width is quantified based on 70~\% quantile of particle positions in radial direction (d70) instead of FWHM. The hybrid CFD/DSMC approach is used along with the molecular drag force model (Eqs.~(\ref{eq:epstein}/\ref{eq:epstein2}) or Eqs.~(\ref{eq:baines_spec}/\ref{eq:baines_diff}) based on the Mach number of the flow). As visible in~\autoref{fig:additionalcases}, the results predicted by the simulation show very good agreement with the experimental data.
This further application case underlines the suitability of the chosen simulation approach.

\section{Other supplementary information}

\subsection{OpenFoam simulation}\label{app:OF}

\begin{figure}[!hbt]
    \centering
    \includegraphics[width=\linewidth]{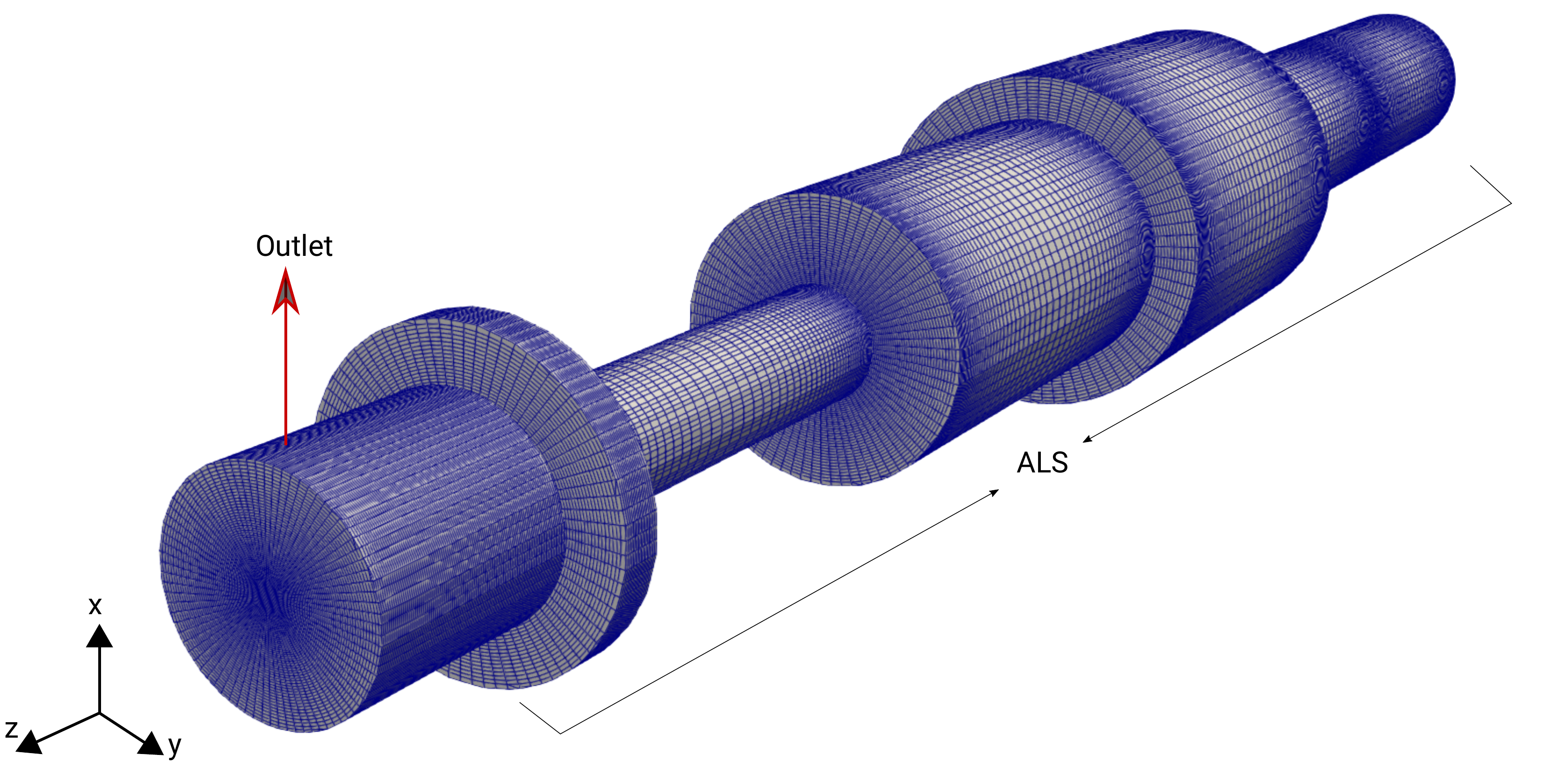}
    \caption{Structured O-grid of the ALS geometry along with the vacuum chamber representation.}
    \label{fig:als_mesh}
\end{figure}

For simulating gas flows through ALS geometries~\cite{Worbs:te5083, worbs_phd2022} using CFD, computational grids are generated using the \texttt{blockMesh} and \texttt{snappyHexMesh} utilities in OpenFoam. Since the ALS geometries are axisymmetric, structured body-fitted standard 3D O-grid type grids are generated as shown in~\autoref{fig:als_mesh}. The vacuum chamber in this simulation is represented by a cylindrical mesh of radius 5~mm and length of 10~mm from the exit of ALS ($z = 0$). Here, the wave transmissive outlet boundary condition is applied~\cite{peravali2024accuracy}. The entire ALS mesh contains a total of approximately $2.44 \times 10^6$ cells, which is based on grid-independence studies. 

\subsection{DSMC simulation}\label{app:dsmc}

The DSMC simulator SPARTA~\cite{SPARTA} uses a Cartesian grid unlike OpenFoam. To resolve the geometry of the ALS properly, and to assure a grid that fits to the entire range of flow Knudsen numbers, a regular grid of size $\Delta x = 5 \times 10^{-5}$~m is used. The time step $\Delta t$ used in the simulation is calculated by $\Delta t = 0.7 \, \Delta x / \overline{v}$ where $\overline{v}= \sqrt{{8 k_B T}/{(\pi \, m)}}$ is the mean thermal speed of the gas molecules. The fully diffusive (isotropic scattering) gas-surface interaction model is used to model the interaction between ALS walls and the gas.  The no-time-counter (NTC) method is employed for collision sampling along with VSS molecular model. The  Larsen and Borgnakke model with constant relaxation is applied to handle the internal energy exchange~\cite{peravali2024accuracy}. The number of DSMC particles per grid cell $N_c \approx 1650$ is used and the number of sampling time steps $N_T$ used were $40,000$ there by giving a sample size $S=N_c \times N_T = 66 \times 10^6$.


%
%

%


\bibliography{aipsamp}

\end{document}